\documentstyle[12pt]{article}
\begin{document}

\title{ On Adomian's Decomposition Method
 for Solving Differential Equations}

\author{P. Di\c t\u a and  N. Grama\\
Institute of Atomic Physics\\
Theoretical Physics Department\\
P.O.Box MG6, Bucharest ROMANIA}

\maketitle
\vskip1.5cm
\begin{abstract}
{We show that with a few modifications the Adomian's method for solving
second order differential equations can be used to obtain the known
results of the special functions of mathematical physics. The modifications
are necessary in order to take corectly into account  the behavior of
the solutions in the neighborhood of the singular points.}
\end{abstract}

\begin{section}{Introduction}
Recently a great deal of interest has been focused on the application of
Adomian's decomposition method to solve a wide variety of stochastic and
deterministic problems \cite{adom}.Although the Adomian's goal is to find a method
to unify linear and nonlinear, ordinary or partial differential equations
for solving initial and boundary value problems, we shall deal in the following
only with linear second order differential equations.

Our aim here is to compare the decomposition method with the classical methods
for solving differential equations in order to obtain a better understanding
of it. Because the special functions are extremely useful tools for obtaining
closed form as well as series solutions to a variety of problems arising in
science and engineering we tryed to reobtain the known results by the new method.
We found that with a slight modification the method works also in this case.

Usually one starts with an equation $ Ff(x)=g(x) $, where $F$ represents a
general nonlinear ordinary differential operator. $F$ is decomposed in two parts
$L+R$, where $L$ is the linear part of $F$ . The core of the method consists
in finding an inverse of $L$ . In many cases this is not possible, in others
the Green function is quite complicate and consequently there are difficulties
of integration. The solution proposed by Adomian \cite{adom} is to take $L$ as
the heighest-ordered derivative of the linear part. 
For example, for initial value problems Adomian defines $L^{-1}$ for $L=d^2/dx^2$
as the two-fold integration operator from $0$ to $x$ . We have $L^{-1}L f(x)=
f(x)-f(0)-xf'(0)$ and therefore
$$f(x)=f(0)+xf'(0) -L^{-1}R +L^{-1}g\eqno1.1$$

Such stated the method is quite old. For linear second order differential 
equations it has been proposed by Cochran, who transformed the equation
$$[p(x)f'(x)]' + q(x)f(x)=g(x)$$
into an equivalent Volterra integral equation \cite{coch}. Cochran has been 
interested 
in
the uniqueness of solutions for two-point boundary-value problems and in
the existence of eigenfunctions of the related homogeneous equation.

Adomian's point of view was to take $L$ as the simplest easily invertible 
operator pushing all the other terms in the reminder $R$ or in the non-
homogeneous part. 
With this choice the method
has given many beautiful results.

Our first comment here is that in order to find the classical results for
the special functions this approach is not sufficient. The second one is that
we give an example, the Bessel polynomials equation, where $L$ has to be
chosen as a {\it first} order differential operator for solving the equation,
although the equation satisfied by them is of second order. The last comment points
out that the approach 1.1 supposes that $f(x)$ and its derivative are
 finite 
at the point $x=0$ , thus
one cannot obtain solutions in the neighborhood of a singular point.

As a first conclusion this means that the decomposition method is a useful tool
for solving concrete problems, but clearly its form cannot be decided 
independent of the problem.
 
We shall consider the linear second order equation written in the more general
form

$$ L(x,D) f(x) - R(x,f,Df,D^2f)  = 0 \eqno 1.2$$
~\\
where $D=d/dx$ and $R$ is the reminder which includes all the other 
terms not contained in the principal part $L(x,D)$ whose form is

$$ L(x,D)= h(x) D p(x) D \eqno1.3$$\\
$h(x)$  and $p(x)$ are smooth functions whose properties are given below.
The minus sign in 1.2 is taken for convenience.

At this stage the clue of the problem is to make a clever choice of $L$
and $R$ such that the resulting pseudo Volterra integral equation should
be easily solvable.

 In many cases bringing 1.2 to the canonical form complicates the matter. We
 propose to find an inverse for the  general form 1.3 and use it to find
many of the known formulas for the solutions of classical special function 
equations.

Our second approach consists in writting 1.2 as a formal non-homogeneous 
equation
and by using a variant of the method of  variation of parameters we 
transform it into a pseudo Volterra integral equation. 
We consider this form as the most
general form of the decomposition method for the second order differential
equations. We apply it to solve the Bessel functions equation.
\end{section}
\begin{section}{Special Functions Equations}
Most of the equations for the classical special functions are of the form 1.2,
with the principal part $L(x,D)$ of the form 1.3. Many of them have $h(x)=1$.

The functions $h(x)$ and $p(x)$ are supposed to be smooth and $1/p(x)$ 
locally
integrable around some point, which we can take to be $x=0$ without loss
of generality. The decomposition method consists in finding an inverse operator
$L^{-1}$ and with its help we are able to construct in many cases the exact
solution of the problem.

A formal inverse of 1.3 can be easily found. We choose it as

$$L^{-1}(x,D) f(x) = \int_0^x  {dt\over p(t)}\int_0^t {
dy\over h(y)} f(y) \eqno2.1$$

One sees that
$$(L L^{-1})f(x) = I\cdot f(x)$$
but $$(L^{-1}L) f(x)\not= I\cdot f(x)$$
where $I$ is the identity operator. The second relation tell us that $L^{-1}$
is not a true inverse, it becomes so when we take into account the initial conditions,
which is equivalent with prescribing the values of $f(0)$ and $f'(0)$ .

Indeed
$$(L^{-1}L)f(x)=\int_0^x{dt\over p(t)}\int_0^t{dy\over h(y)}h(y)
{d\over dy} p(y){df(y)\over dy}$$
$$=\int_0^x{dt\over p(t)}[p(y){df\over dy}]_0^t =\int_0^x{df(t)\over dt}
-p(0)f'(0)\int_0^x {dt\over p(t)}$$
$$= f(x) -f(0) -p(0)f'(0)\int_0^x{dt\over p(t)}$$
Thus we obtain from 1.2
$$f(x)=f(0)-p(0)f'(0)\int_0^x{dt\over p(t)}+\int_0^x {dt\over p(t)}\int_0^t
{dy\over h(y)}R(y,Df) \eqno2.2$$
which is a Volterra integral equation.. Its most general form 
for second order differential equations will be given 
in Section 3. The solution of the last equation is sought as a series
$$f(x)=\sum_{k=0}^{\infty}f_k(x)$$
 where $$f_0=f(0)-p(0)f'(0)\int_0^x {dt\over p(t)}$$
and $f_k(x)$ is obtained by the Picard method of succesive approximation
$$f_k=L^{-1}(x,D)f_{k-1}$$

In the following we shall apply the above technique to a few classical
equations.
%
\begin{subsection} {Gauss Hypergeometric Function}

Because the solutions to many mathematical physics problems are expressed
by the hypergeometric functions we consider first this equation whose
standard form is
$$x(1-x)f''(x) + [\gamma -(\alpha+\beta+1)x]f'(x)-\alpha\beta f(x)=0$$
and rewrite it as
$$Lf(x)= xf''(x)+\gamma f'(x)\equiv x^{1-\gamma}{d\over dx}(x^\gamma{df\over dx})$$
$$=
x^2 f''(x)+(\alpha+\beta+1)xf'(x)+\alpha\beta f(x)$$

In this case $h(x)=x^{1-\gamma}$ and $p(x)=x^{\gamma}$. We suppose that $\gamma>0$
and then
$$ (L^{-1}f)(x)= f(x)-f(0)$$

We take as usual in the decomposition method
$$f(x)= f_o(x)+f_1(x)+f_2(x)+\dots$$
and choose the initial condition as
$$f_0=f(0)=1$$
Using the form 2.2 we get the recurrence relation
$$f_n =L^{-1}[x^2f''_{n-1}(x)+(\alpha+\beta+1)xf'_{n-1}(x)+\alpha\beta f_{n-1}(x)]$$
$$=\int_0^x x^{-\gamma}dx\int_0^x x^{\gamma-1}[x^2f''_{n-1}(x)+(\alpha+\beta+1)xf'_{n-1}(x)+\alpha\beta f_{n-1}(x)]dx$$
We find that
$$f_{1}
 = {\alpha \beta\over
\gamma}{x\over 1!
}$$
and by induction 
$$ f_n ={(\alpha)_n  (\beta)_n \over{(\gamma)_n}}{x^n\over n!}$$
where $(\alpha)_n=\alpha(\alpha+1)\dots (\alpha +n-1)$

In this way we get the well known solution
$$f(x)=\,_2F_1(\alpha,\beta;\gamma;x)=1+\sum_{n=1}^{\infty}{(\alpha)_n(\beta)_n\over (\gamma)_n}{x^n\over
n!}$$
\end{subsection}
\begin{subsection}{Confluent Hypergeometric Functions}
As concerns the confluent hypergeometric equation
$$xf''(x)+(\gamma-x)f'(x)-\alpha f(x)=0$$
its solution can be obtained in the same manner as that of the Gauss 
hypergeometric function, the functions $h(x)$ and $p(x)$ being the same.

We shall consider now the equation
$$f''(x)+ax^{q-1}f'(x)+b x^{q-2} f(x)=0 \eqno2.3$$
where $a,b$ and $q$ are complex numbers, whose solution seems to be unknown.
 For $q=0$ it reduces to Euler equation.
For others values of $q$ it is equivalent with a degenerate hypergeometric function.
More about this equation see \cite[eq. 2.60]{kam}.

We construct two independent solutions in the neighborhood of the point $x=0$.
We take $L=D^2$ and then
$$L^{-1}f(x)=f(x)-f(0) -xf'(0)$$
The first solution is defined by
$$f(0)=1 ,~~~~~~~  f'(0)=0\eqno2.4$$
We choose $$f_0=f(0)=1$$
and find
$$f_1=-{b\over q(q-1)}x^q\eqno2.5$$
We make the ansatz
$$f_k=c_k x^{kq}$$
and from the relation $f_{k+1}=L^{-1}f_k$ we get
$$f_{k+1}=-c_k \int_0^x dx\int_0^x [aqkx^{(k+1)q-2}+bx^{(k+1)q-2}]$$
$$=
-c_k {aqk+b\over (k+1)q-1}{x^{(k+1)q}\over q(k+1)}$$

In this way we find the recurrence relation
$$c_{k+1} =-{aqk+b\over{[q(k+1)-1]q(k+1)}}c_k$$
Taking into account 2.5 we find

$$c_n = (-1)^n {b(aq+b)\dots((n-1)aq+b)\over{(q-1)(2q-1)\dots(nq-1)}}{1\over
q^n  n!}$$
$$=(-1)^n{(b/aq)_n\over (1-{1/q})_n}{\biggl({a\over q}\biggr)^n} {1\over n!}$$

The solution of the Eq. 2.3 with the initial conditions 2.4 is
$$f(x)=1+\sum_{n=1}^{\infty} (-1)^n {\bigl(b/aq\bigr)_n\over(1-{1/q})_n}
{\biggl({a\over q}\biggr)^n}
{x^{nq}\over n!}$$
i.e. $f(x)$ is the confluent hypergeometric function
$$f(x)=\, _1F_1 ({b\over aq},1-{1\over q};-{ax^q\over q})\eqno2.6$$
The second solution which satisfies
$$f(0)=0,~~~~~f'(0)=1$$
is found in the same way and is given by
$$f(x)=x\,_1F_1({a+b\over aq},{q+1\over q};-{ax^q\over q})\eqno2.7$$

The solutions 2.6 and 2.7 give the general solution of Eq.2.3 and as function
of $q$ have poles at $q=\pm1/n+1$,respectively, which accumulate
at $q=0$. When $b/aq=-p$, and $a+b/aq=-n$, respectively the solutions reduce
to polynomials in $x^q$.
\end{subsection}

\begin{subsection}{Bessel Polynomials}
Until now in all the applications of the decomposition method the operator $L$
contained a piece involving the highest-ordered derivative. We want to
present an example where such an approach does not work. If we form $L$ 
from lower-ordered derivatives we can solve the equation.

We consider the equation
$$x^2 f''(x) +(2 \alpha x+\beta)f'(x)-\gamma(\gamma+1)f(x)=0\eqno2.8$$
where $\alpha$, $\beta$ and $\gamma$ are complex numbers. 
If $\alpha=1, ~~\beta=2$ and $\gamma=n$ the above equation is the defining equation
for the Bessel polynomials \cite{kral} \cite{gro}.

We take $L$ as the first order operator
$$L={d\over dx}$$
The above equation is rewritten as
$$Lf(x)=f'(x)={1\over\beta}[\gamma(\gamma+1)f(x)-2x\alpha f'(x)-x^2 f''(x)]$$
In this way we find the Volterra equation 
$$f(x)=f(0) + {1\over\beta}\int_0^x [\gamma(\gamma+1)f(x)-2x\alpha f'(x)-x^2 f''(x)]\,dx$$
We look for a regular solution of Eq. 2.8 at the origin and take 
  the initial condition   
$$f_0=f(0)=1$$
We get
$$f_1={\gamma(\gamma+1)\over\beta}{x\over 1!}$$
We make the ansatz
$$f_k={c_k \over \beta^k} {x^k\over k! }$$
From the integral equation we find 
$$f_{k+1}=-{c_k [k^2+(2\alpha-1)k-\gamma(\gamma+1)]\over(k+1)!}{x^{k+1}\over
  \beta^{k+1}
}$$
Now we define
$$(2\nu+1)^2=(2\alpha -1)^2 +4\gamma(\gamma+1)$$
and from the above relations
one finds the recurrence relation
$$c_{k+1}=-(k+\alpha +\nu)(k+\alpha -\nu-1)c_k $$
whose solution is
$$c_k=(-1)^k (\alpha+\nu)_k (\alpha-\nu-1)_k$$
and the solution of Eq. 2.7 is
$$f(x)=\,_2F_0(\alpha-\nu-1,\alpha+\nu;-{x\over\beta})$$

For $\alpha-\nu-1=-n$ we find the generalized Bessel polynomials [4].
\end{subsection}

\end{section}
\begin{section}{Volterra Integral Equation}

Let be a general second-order differential equation
$$f''(x)+a(x)f'(x)+b(x)f(x)=h(x) \eqno3.1$$
and let be $\varphi(x)\not=0$ a solution of the corresponding homogeneous
equation.

By applying a variant of the method of variation of parameters  we
obtain the general solution of Eq. 3.1 as
$$f(x)=C_1\varphi(x)+C_2\varphi(x)\int{dx\over E(x)\varphi^2(x)}+$$
$$
\varphi(x)\int{dx\over E(x)\varphi^2(x)}\int E(x)\varphi(x)h(x)dx \eqno3.2$$
where $E(x)=exp\int a(x)dx$ and $C_1$ , $C_2$ are constants \cite{kam}.

In the above form $L^{-1}$ is defined as an indefinite integral; for every 
problem we have to transform it into a definite integral according with
the solution we are looking for.

We consider 3.2 as the starting point in the formulation of the 
most general form of the decomposition method.
We use Eq. 3.1 to transform a linear homogeneous equation into a pseudo
Volterra integral
equation. The crucial point is to separate a convenient part in the left hand side
of 3.1 such that we can easily find a solution of this part; 
this separation has to take into account the behavior of the solution in the 
neighborhood of the point where we are looking for the solution, to improve our
chance to obtain the full series in explicit form.
The simplest separation is  
by moving the term $b(x)f(x)$ on the right-hand side, then a solution of the left
side alone is $f(x)=1$ . We shall give a few examples rather than treat the general
case.

Let us come back to the hypergeometric equation 2.1 and write it as
$$f''(x)+{\gamma\over x}f'(x)=xf''(x)+(\alpha+\beta+1)f'(x) +{\alpha\beta
\over x}f(x)$$
We treat the right-hand side of this equation as a formal non-homogeneous term
and look for a solution of the homogeneous part, i.e. of the left part.
The  solutions are $\varphi _+(x)=1$  and $\varphi _-(x)=x^{1-\gamma}$
We consider first $\varphi _+$ and take 
$C_1=1$ and $C_2=0$ . 
Since 
we 
look for the regular solution 
 at $x=0$ we take $L^{-1}$ as a 2-fold integration
from $0$ to $x$ and find from
3.2 the Volterra integral equation
$$f(x) =1+\int_0^x{dx\over x^{\gamma}}\int_0^xx^{\gamma}[xf''(x)+(\alpha+\beta+1)f'(x)+{\alpha\beta\over x}f(x)] dx\eqno3.3$$
We solve Eq. 3.3 by iteration taking  the zero-order approximation the non-
homogeneous term
$f(0)=f_0=1$
and find
$$f_{n+1}=\int_0^x{dx\over x^{\gamma}}\int_0^x x^{\gamma} [xf''_n(x)+(\alpha+\beta+1)f'_n(x) +{\alpha\beta\over x}f_n(x)]dx$$
which coincides with the result found in the preceding section.

If we start with the second solution $\varphi_-$ and take as above $C_1=1$ and
$C_2=0$ we find the integral equation

$$f(x) =x^{1-\gamma}+ x^{1-\gamma} \int_0^x{dx\, x^{\gamma-2}}\int_0^x[xf''(x)+(\alpha+\beta+1)f'(x)+{\alpha\beta\over x}f(x)] dx$$
We solve it by iteration, but now the zero-order approximation  is $f_0=x^{1-\gamma}$,
and find
$$f_1=x^{1-\gamma}{(\alpha-\gamma+1)(\beta-\gamma+1)\over{ 2-\gamma}}{x\over 1!}$$

By induction we get the series for the second solution
$$f(x)=x^{1-\gamma}\,_2F_1(\alpha-\gamma+1,\beta-\gamma+1;2-\gamma;x)$$

We consider now the equation of the Bessel functions and write it as

$$f''(x)+{1\over x}f'(x)+\bigl(1-{\nu^2\over x^2}\bigr)f(x)=0$$

Its solution cannot be obtained by the decomposition method as originally
stated, nor in the more general form of it given by us in Sec. 2. 
The reason is the singular behavior of the Bessel functions 
in the neighborhood of  the point
$x=0$;
however its
solution can be easily obtained using the form 3.2 . Indeed,  we rewrite
the Bessel equation in the form
$$f''(x)+{1\over x}f'(x)-{\nu^2\over x^2}f(x)=-f(x)$$
and observe that the "homogeneous" part has the solutions
$\varphi_{\pm}(x)=x^{\pm\nu}$ . Since the behaviour of the Bessel functions
at $x=0$ is that given by $\varphi_{\pm}$ we choose $L^{-1}$ as a 2-fold
integration from $0$ to $x$ .

 Let us try the first solution; we choose
$C_1=2^{-\nu}/\Gamma(\nu+1)$ and $C_2=0$ and from 3.2 we get the integral equation

$$f(x)={(x/2)^{\nu}\over\Gamma(\nu+1)} -x^{\nu}\int_0^x dx x^{-1-2\nu}\int_0^x
dx x^{1+\nu}f(x)$$
The initial approximation of the solution is 
the non-homogeneous term which is no more constant
$$f_0={(x/2)^{\nu}\over\Gamma(\nu+1)}$$
and from the above equation we get
$$f_1=-{(x/2)^{\nu+2}\over\Gamma(\nu+2)}$$
and by induction we find the known series.
$$f(x)=J_{\nu}(x)=\sum_{k=0}^{\infty} (-1)^k{(x/2)^{\nu+2k}\over k!\Gamma(\nu+k+1)}$$

If  we choose the initial conditions as
$$C_1=0,~~~~C_2={2^{-\nu}\over\Gamma(1-\nu)}$$  proceeding as above we find the
second independent solution, $J_{-\nu}(x)$ .

The equation 3.2 represents the starting point for obtaining 
the most general form of the decomposition method.
Indeed it can be recovered, in the form given by Adamian, by the choice
$a(x)=b(x)=0$ in Eq. 3.1. As a consequence $\varphi(x)=1$ and $E(x)\equiv1$
and if we look for regular solutions in the neighborhood of $x=0$
 3.2 has the form

$$f(x)=C_1+C_2x +\int_0^x dx\int_0^x h(x)dx$$
and by identification $$C_1= f(0)~~~C_2=f'(0)$$
\end{section}

\begin{section}{Conclusion}
We have shown that with the necessary modificatons the Adomian's method
can be used to obtain the classical results on the special functions.
Rather than prescribe a unique form we show that the concrete problem decides how
to formulate the method.
The clue of this one consists in transforming a linear differential equation into
a pseudo Volterra integral equation whose solution is obtained by the Picard
process of succesive approximation.

The method can be obtained by a light modification of the known form of the
solution for a linear second order differential equation, the sole difference
being the extension of what is ussualy called
the non-homogeneous term in the sense that it can include important pieces
from the homogeneous part.

Our results can also be seen  as a good  illustration for the effectivness of
 the Picard
method of succesive approximation.
\end{section}

\end{document}